\documentclass[runningheads]{llncs}
\usepackage{graphicx}
\usepackage{svg}
\usepackage{hyperref}
\usepackage{caption}
\usepackage{subcaption}
\usepackage{amssymb}
\usepackage{floatrow} 
\newfloatcommand{capbtabbox}{table}[][\FBwidth]

\begin{document}
\title{Structure-Preserving Multi-Domain Stain Color Augmentation using Style-Transfer with Disentangled Representations}  
\titlerunning{Structure-Preserving Multi-Domain Stain Color Augmentation}
%
\author{Sophia J. Wagner \inst{1,2,4}, Nadieh Khalili \inst{5}, Raghav Sharma \inst{3}, Melanie Boxberg \inst{1,4}, Carsten Marr \inst{3}, Walter de Back \inst{5}, Tingying Peng \inst{1,2,4}}

%
\authorrunning{S. J. Wagner et al.}
%
\institute{Technical University Munich, Munich, Germany \and 
Helmholtz AI, Neuherberg, Germany \and 
Institute for Computational Biology, HelmholtzZentrum Munich, Germany \and
Munich School of Data Science (MuDS), Munich, Germany \and
ContextVision AB, Stockholm, Sweden}
%
\maketitle              
\begin{abstract}
In digital pathology, different staining procedures and scanners cause substantial color variations in whole-slide images (WSIs), especially across different laboratories. These color shifts result in a poor generalization of deep learning-based methods from the training domain to external pathology data. To increase test performance, stain normalization techniques are used to reduce the variance between training and test domain. Alternatively, color augmentation can be applied during training leading to a more robust model without the extra step of color normalization at test time. We propose a novel color augmentation technique, HistAuGAN, that can simulate a wide variety of realistic histology stain colors, thus making neural networks stain-invariant when applied during training. Based on a generative adversarial network (GAN) for image-to-image translation, our model disentangles the content of the image, i.e., the morphological tissue structure, from the stain color attributes. It can be trained on multiple domains and, therefore, learns to cover different stain colors as well as other domain-specific variations introduced in the slide preparation and imaging process. We demonstrate that HistAuGAN outperforms conventional color augmentation techniques on a classification task on the publicly available dataset Camelyon17 and show that it is able to mitigate present batch effects. \footnote[1]{Code and model weights are available at \url{https://github.com/sophiajw/HistAuGAN}.}

\end{abstract}

\section{Introduction}

Modern cancer diagnosis relies on the expert analysis of tumor specimens and biopsies. To highlight its structure and morphological properties, conventionally, the tissue is stained with hematoxylin and eosin (H\&E) \cite{Chan2014-ns}. The path from the raw tissue to the final digitized image slide however consists of many different processing steps that can introduce variances, such as tissue fixation duration, the age and the composition of the H\&E-staining, or scanner settings. Therefore, histological images show a large variety of colors, not only differing between laboratories but also within one laboratory \cite{Bejnordi2016-eb}.

This variability can lead to poor generalization of algorithms that are trained on WSIs from a single source. One strategy to account for this is stain color normalization. 
Traditionally, this is either done by aligning the color distribution of the test images to a reference tile in the training domain \cite{Reinhard2001-op} or by decomposing the color space of a reference tile into hematoxylin and eosin components \cite{macenko2009method,Vahadane2016-lu}. Then, H\&E components of the test tiles can be aligned while keeping the structure intact.

Recently, the focus shifted toward the application of style-transfer methods such as cycle-consistent generative adversarial networks, CycleGAN \cite{Zhu2017-li}, for stain normalization \cite{Tschuchnig2020-tj}. However, these models aim to match the target distribution possibly leading to undesired changes in the morphological structure \cite{hallucinatingGANs}. 
To circumvent this, other approaches propose color space transformations \cite{Shaban2019-sj}, structural similarity loss functions \cite{Liang2020-fl}, or residual learning \cite{De_Bel2021-ay}.

We propose a novel histological color transfer model, HistAuGAN, based on a GAN architecture for image-to-image translation. In contrast to previous approaches, HistAuGAN disentangles the content of a histological image, i.e., the morphological tissue structure, from the stain color attributes, hence preserving the structure while altering the color. Therefore, HistAuGAN can be used as a stain augmentation technique during training of a task-specific convolutional neural network (CNN). We demonstrate that this helps to render the trained network color-invariant and makes it transferable to external datasets without an extra normalization step at test time. Applied as an augmentation technique, HistAuGAN significantly outperforms other color augmentation techniques on a binary tumor-classification task. Furthermore, clustering results suggest that HistAuGAN can capture sources of domain shifts beyond color variations, such as noise and artifacts introduced in the staining or digitization process, e.g., image compression or blurring.

To the best of our knowledge, HistAuGAN is the first GAN-based color augmentation technique that generates realistic histological color variations.   

\section{Method}
\subsection{Model architecture}

\begin{figure}[t]
    \centering
    \includegraphics[width=0.9\textwidth]{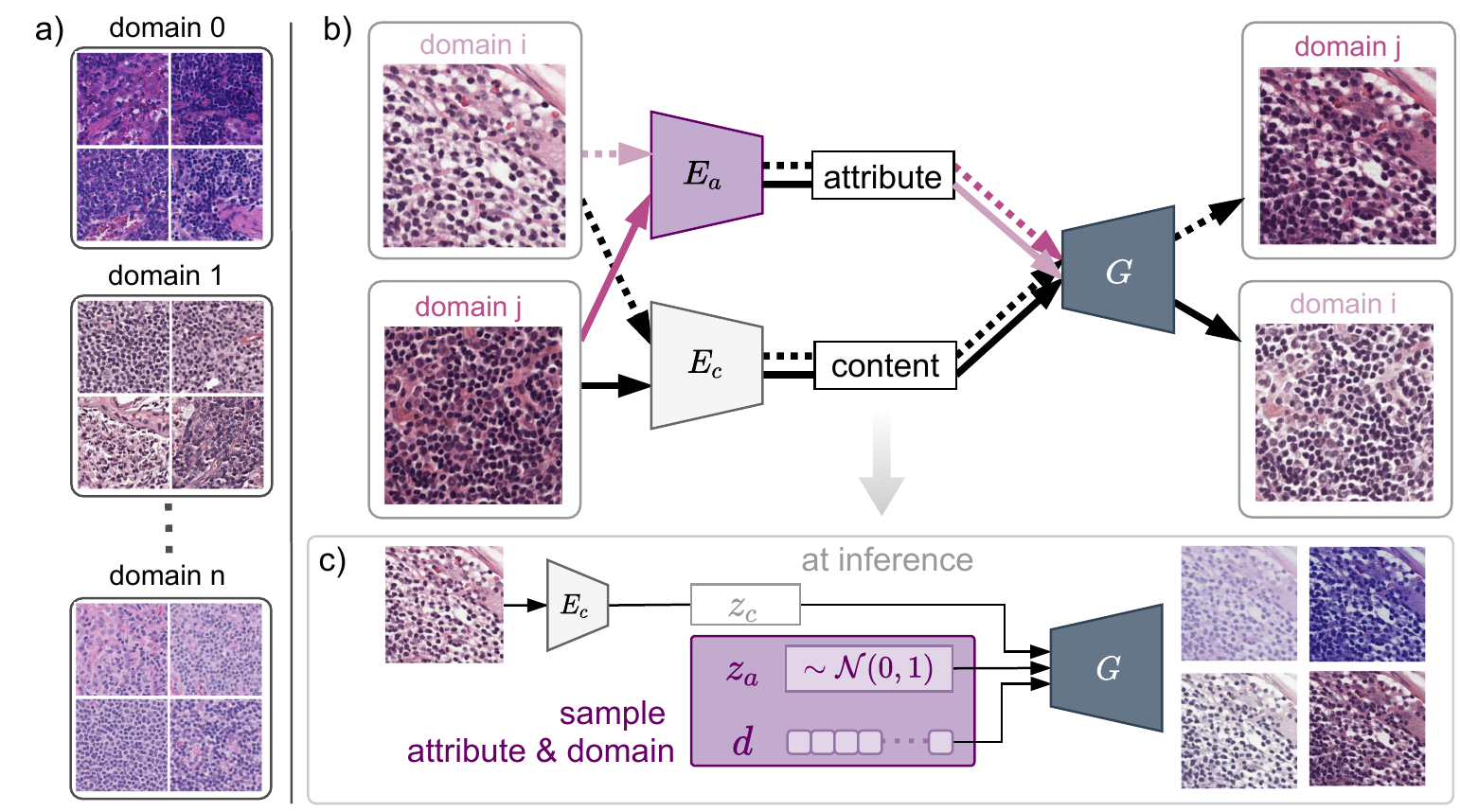}    
    \caption{We propose HistAuGAN for structure-preserving multi-domain stain color augmentation. (a) Histological slides from different laboratories (domains) exhibit color variations. (b) Model architecture. Here, the domain information flow is visualized by colored arrows. (c) At inference, HistAuGAN can be used as an augmentation technique by sampling attribute $z_a$ and domain $d$.}
    \label{fig:model}
\end{figure}

We build our model based on a multi-domain GAN using disentangled representations, inspired by DRIT++ \cite{Lee2020-zw}. Originally designed for image-to-image translation of natural images using a predefined style, we propose its application on histological images to disentangle the morphological tissue structure from the visual appearance. In contrast to previous CycleGAN-based color normalization methods that use only a single encoder, HistAuGAN is able to separate two essential image properties from each other as visualized in Figure \ref{fig:model}b: the domain-invariant content encoder $E_c$ encodes the histopathological structure of the tissue, e.g., size and position of the nuclei, whereas the domain-specific attribute encoder $E_a$ learns the domain-specific color appearance. The model can be trained on data from multiple domains and thereby captures both inter-laboratory variability between multiple domains and intra-laboratory variability within each domain at the same time. Finally, the generator $G$ takes as input a content vector $z_c$, an attribute vector $z_a$, and the one-hot-encoded domain vector $d$ and outputs a simulated histological image. The objective function is given by 

\begin{equation}
    L_{total}=w_{cc}L^{cc}+ w_cL^c + w_dL^d + w_{recon}L^{recon}+w_{latent}L^{latent}+w_{KL}L^{KL},
\end{equation}

where $L^{cc}$ is the cycle-consistency loss, $L^c$ and $L^d$ are adversarial losses for the content and the attribute encoder, $L^{recon}$ is an $L_1$-loss for image reconstruction, $L^{latent}$ is an $L_1$-loss for latent space reconstruction, and $L^{KL}$ enforces the latent attribute space to be distributed according to the standard normal distribution. Please refer to \cite{Lee2020-zw} for a detailed explanation of each loss and the precise hyperparameter setting. 

\begin{figure}[ht]
    \centering
    \includegraphics[width=0.7\textwidth]{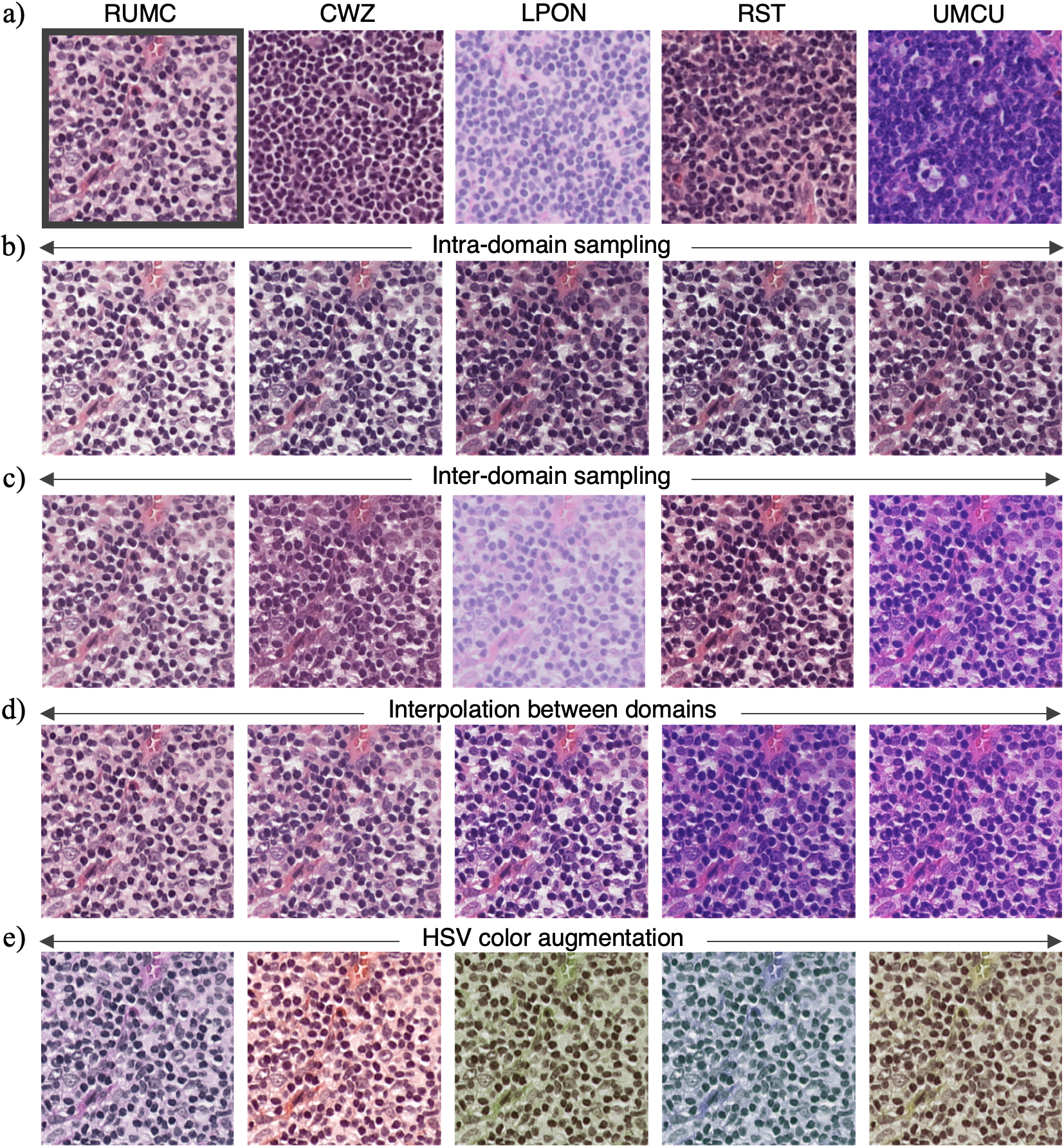}
    \caption{Overview of the color variation in the dataset and the augmentation techniques used in this paper using the framed image as example tile.}
    \label{fig:data}
\end{figure}

At inference, using the fixed content encoding of the input image $z_c$, we can sample the attribute vector $z_a$ and the one-hot encoded domain vector $d$ as visualized in Figure \ref{fig:model}c. Hence, we can map one image to many different structure-preserving augmentations. More specifically, we sample a random color attribute $z_a$ from a normal distribution that parametrizes the stain color variabilities in one domain. Figure \ref{fig:data}b shows randomly sampled outcomes of intra-domain augmentations. Additionally, we can change the one-hot-encoded domain vector $d$ to project the input image into multiple target domains as visualized in Figure \ref{fig:data}c. In addition to sampling from the training domains, we can also interpolate between these domains to obtain an even broader variety of realistic color appearances for histopathological images. Figure \ref{fig:data}d demonstrates this by linearly interpolating the domain from domain RUMC to domain UMCU according to 
\begin{equation}
    d = (1-t) \cdot d_{\mathrm{RUMC}} + t \cdot d_{\mathrm{UMCU}}, \quad \mathrm{for } \  t \in [0,1].
\end{equation}

\subsection{Competing methods for stain color augmentation}

Most existing stain color transfer methods are used for stain normalization, i.e., to transfer the stain color of the test domain to that of the training domain. 
Recently, it has been shown that simple stain color augmentations, such as perturbing the HSV color space of the histological images, perform better and lead to more robust models than traditional and network-based normalization techniques \cite{Tellez2019-nt}. Therefore, we compare our HistAuGAN to the HSV augmentations used in \cite{Tellez2019-nt}. Besides HSV augmentation, there is a more complicated augmentation technique based on the Wasserstein distance of different domains \cite{Nadeem2020-bn}. But the method is much slower than HSV and HistAuGAN, thus difficult to be used as an on-the-fly augmentation technique. 

For a quantitative evaluation of our augmentation technique, we consider the following augmentation methods:
\begin{itemize}
    \item \textit{Geometric augmentations}: vertical and horizontal flipping, as well as $90^\circ$, $180^\circ$, and $270^\circ$ rotations.
    \item \textit{HSV color augmentations}: geometric augmentations with Gaussian blur and contrast and brightness perturbations applied with probability 0.25 and 0.5, respectively. We tried both light and strong color augmentations, as suggested in \cite{Tellez2019-nt}. Strong color augmentations can generate unrealistic color appearances. However, applying hue and saturation jittering with factor 0.5 and probability 0.5, which results in relatively strong color perturbance as shown in Figure \ref{fig:data}e, performed best for us.
    \item \textit{HistAuGAN}: geometric augmentations combined with our augmentation technique applied to half of the images during training. For each image, we randomly pick a target domain from the training domains and sample a color attribute vector $z_a \in \mathbb{R}^8$ from the standard normal distribution. 
\end{itemize}

\subsection{Evaluation}

We evaluate HistAuGAN on three different aspects, in particular, i) whether it can remove batch effects present in histological images collected from multiple medical laboratories, ii) how it affects the out-of-domain generalization of a deep learning model trained for a specific down-stream task, and
iii) how HistAuGAN preserves morphological structure during augmentation. 
For ii), we choose a binary classification task of classifying WSI tiles into the classes \textit{tumor} versus \textit{non-tumor} as described in more detail in Section \ref{sec:quant-results}. Question iii) is evaluated by asking a pathology expert to check image similarity before and after augmentation. To explore how generalizable our model is, we extend the HistAuGAN training data (lymph nodes) by tiles from unseen tissue and tumor types, in particular, breast tissue \cite{mitosisdataset}.

\section{Results and Discussion}

\subsection{Dataset}
For the quantitative evaluation of HistAuGAN, we choose the publicly available Camelyon17 dataset \cite{Bandi2019-bo} that provides WSIs from five different medical centers (denoted by RUMC, CWZ, UMCU, RST, and LPON) with different scanning properties and stain colors as shown in Figure \ref{fig:data}a. Pixel-wise annotations are given for 50 WSIs in total, 10 from each medical center. To create the training patches, we first threshold the images with naive RGB thresholding combined with Otsu thresholding and then patch the tissue regions of each WSI at the highest resolution based on a grid into tiles of size $512\times512$ pixels. Each tile is labeled as \textit{tumor} if the ratio of pixels annotated as tumor pixels is larger than 1\%, otherwise, it is labeled as \textit{non-tumor}. The tiled dataset has an \textit{imbalanced} class distribution, i.e., overall, 7\% of the tiles are labeled as \textit{tumor} and the ratio of tumor tiles is in the same order of magnitude across all medical centers.

\subsection{Evaluation of batch-effect removal}
\begin{figure}[t]
    \centering
    \includegraphics[width=0.85\textwidth]{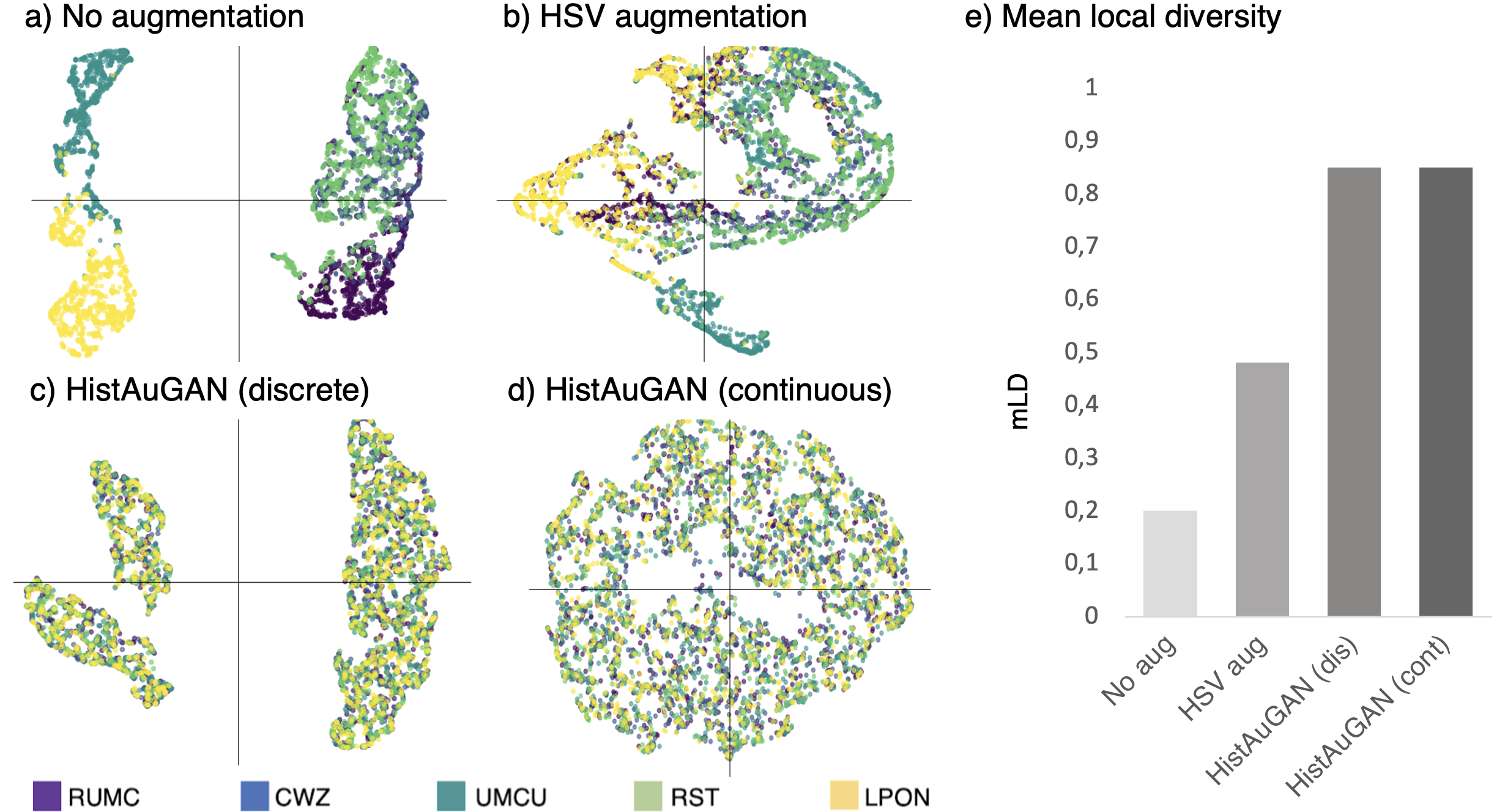}
    \caption{Effect of color augmentation on batch effects in color statistics. (a-d) UMAP embeddings of color statistics of training data, color-coded by source domains. (e) The quantification of mixing based on mean local diversity (mLD, higher is better) suggests HistAuGAN effectively mitigates batch effects.}
    \label{fig:clustering}
\end{figure}
To evaluate how color augmentation mitigates batch effects, we quantify the mixing of images from different medical centers with respect to their color statistics. A random set of 1,000 image tiles were extracted from the WSIs from each center and analyzed in terms of the average values of each component after transformation to various color spaces (RGB, HSV, LAB, HED, grayscale). To visually observe batch effects, we reduced the dimensionality to 2D using UMAP \cite{Becht2018-hk} and labeled points according to their domain as shown in Figure \ref{fig:clustering}a-d. To quantify the mixing of different domains, we measured the mean over the local diversity (mLD) for all $k$-nearest neighborhoods ($k=10$) in the 2D projection using Shannon's equitability which varies between 0 for non-mixed and 1 for perfectly mixed populations (cf. Figure \ref{fig:clustering}e).

Without color augmentation, we observe a clear batch effect: tiles from different domains form distinct clusters ($\mathrm{mLD}=0.2$, Figure \ref{fig:clustering}a). HSV augmentations improve data mixing, but domain-correlated clusters are still visible ($\mathrm{mLD}=0.48$, Figure \ref{fig:clustering}b) and single domains, e.g. LPON, are not mixed with other domains. In contrast, HistAuGAN mixes data from multiple domains (Figure \ref{fig:clustering}c,d) with a high local diversity ($\mathrm{mLD}=0.85$). If HistAuGAN is used to transfer colors to discrete domains, the distinct domain clusters are retained, but each cluster contains well-mixed image samples transferred from all domains (Figure \ref{fig:clustering}c). When HistAuGAN is used to randomly interpolate between domains, a continuous well-mixed color subspace is obtained without any clustering structure (Figure \ref{fig:clustering}d).

These results show that HistAuGAN is highly effective in removing batch effects present in color statistics of images sampled from different medical centers. 

\subsection{Evaluation on a down-stream classification task} \label{sec:quant-results}
\begin{figure}[tb]
    \centering
    \begin{subfigure}[b]{0.49\textwidth}
        \centering
        \includegraphics[width=\textwidth]{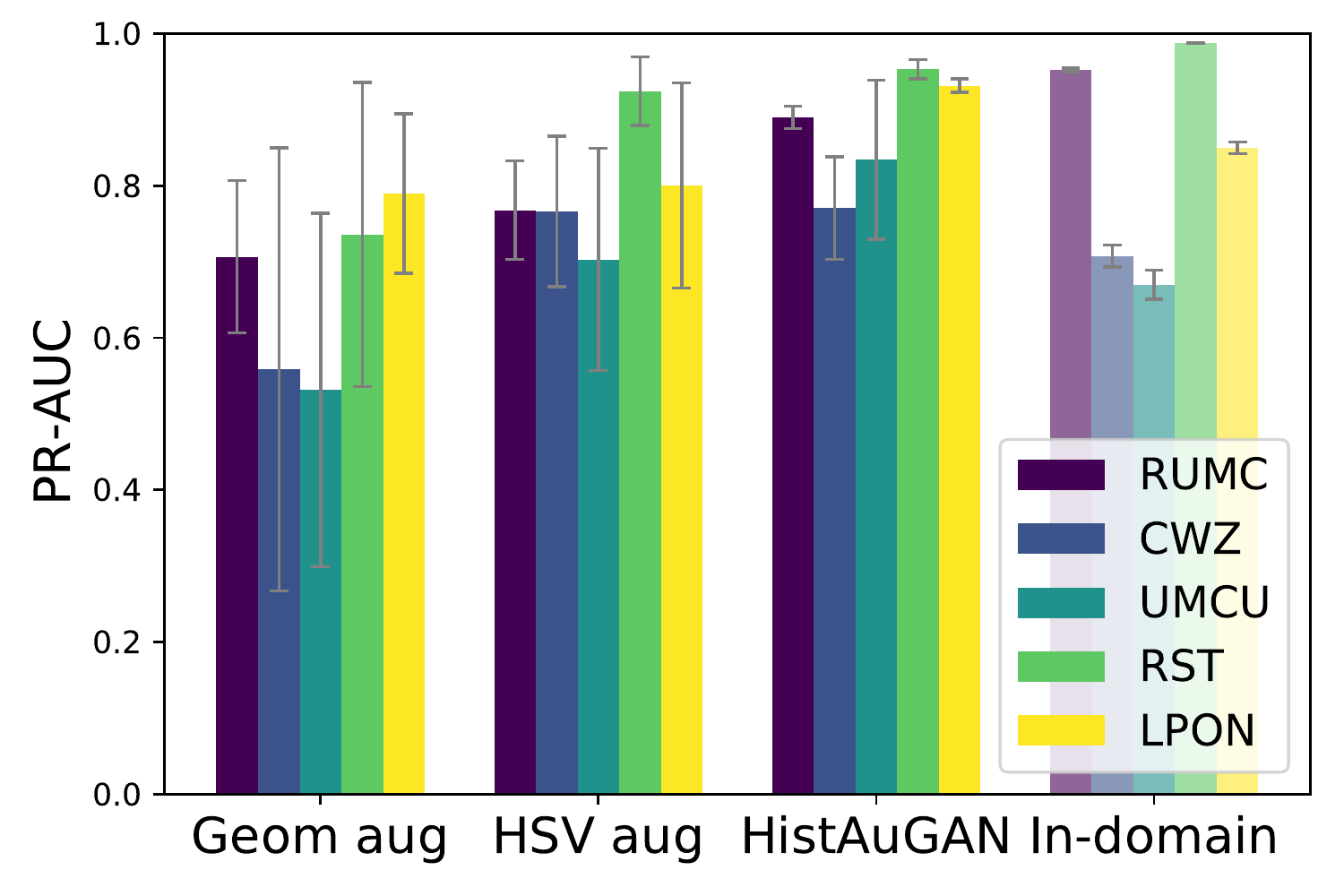}
    \end{subfigure}
    \hfill
    \begin{subfigure}[b]{0.49\textwidth}
        \centering
        \includegraphics[width=\textwidth]{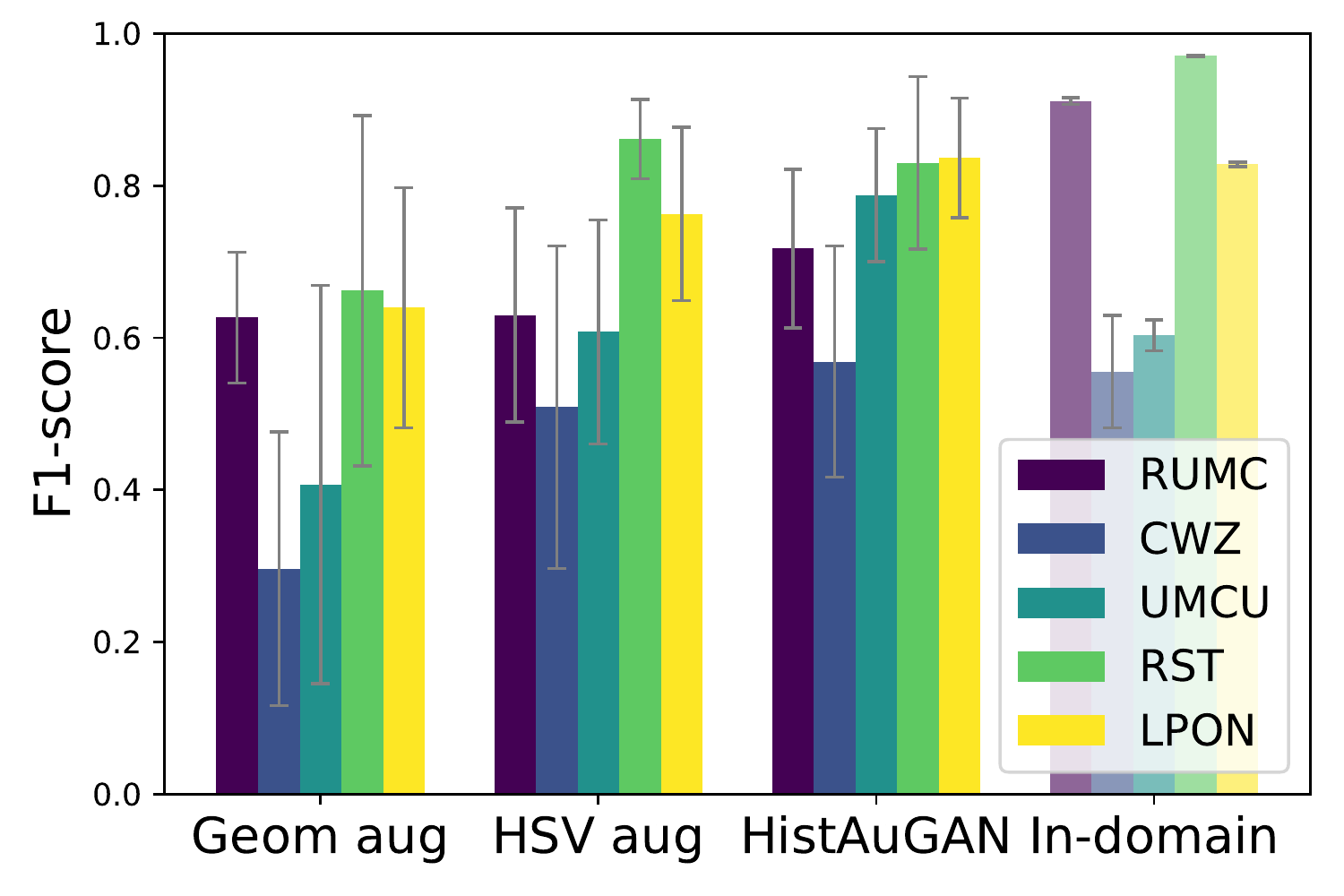}
    \end{subfigure}
    \caption{Precision-recall AUC (left) and F1-score (right) of our binary classification task. The bold bars depict the results on the out-of-domain centers averaged across all runs. The most-right, pale bars denote the in-domain test performance of the classifiers trained with geometric augmentations.}
    \label{fig:quant-results}
\end{figure}

To evaluate the effect of our proposed augmentation method, we train a CNN on a binary tumor classification task and compare the performance on different out-of-domain test sets based on the Camelyon17 dataset.
Due to the relatively small size of our dataset, in particular the small number of tumor tiles, we choose a small CNN, namely, a pre-trained ResNet18 \cite{resnet}, and fine-tune the last two ResNet-blocks together with the fully-connected layer on our dataset. For training, we use weighted cross-entropy-loss to rebalance the contribution of each class, with a learning rate of 1e-5 and an $L_2$-regularization of 1e-5 across all runs and for all augmentation techniques. Furthermore, we used random erasing as regularization on all augmentation techniques \cite{randomerasing}. Since our dataset is highly imbalanced, we report the F1-score of the tumor class in addition to the area under the precision-recall curve (PR-AUC). 

Figure \ref{fig:quant-results} shows the results of the quantitative evaluation of different augmentation techniques on the binary tumor-classification task. For each medical center, we trained three classifiers, one for each augmentation type, and aggregated the results evaluated on the test domains. All experiments were repeated three times. On both metrics, HistAuGAN shows better performance on all of the out-of-domain test sets. As visualized in Figure \ref{fig:data}, the appearance of images from medical center UMCU and LPON deviates strongly from the other centers, explaining their lower scores. In comparison to HSV color augmentation, HistAuGAN performs better in handling the stain color discrepancy between training and test domain and is therefore able to generate a more robust classification model that generalizes better to out-of-domain test sets. This can also be measured in the standard deviation of the results across the out-of-domain test sets centers. For our model, the standard deviation of the PR-AUC for the tumor class amounts to 0.08, whereas it higher for geometric (0.22) and color (0.14) augmentations, respectively, which demonstrates that our model is more robust to underlying stain color variations. The right-most group shows the in-domain test results for geometric augmentations. It can be seen as an upper bound for any stain normalization technique, and thus shows that HistAuGAN can even outperform stain normalization techniques on some of the five domains.

\subsection{Qualitative evaluation by an expert pathologist}
\begin{figure}[tb]
\begin{floatrow}%
\capbtabbox{%
  \caption{Expert evaluation.}%
  \label{tab:expert}%
  }{%
    \begin{tabular}{lcccc} \hline
        Tissue type & High  & Moderate  & Low   & Total \\ \hline
        Lymph nodes & 10    & 7         & 3     & 20    \\
        Breast      & 14    & 4         & 2     & 20    \\ \hline
    \end{tabular}
}
\ffigbox{%
  \caption{HistAuGAN on unseen tissue.}%
  \label{fig:breast}%
}{%
    \includegraphics[width=0.4\textwidth]{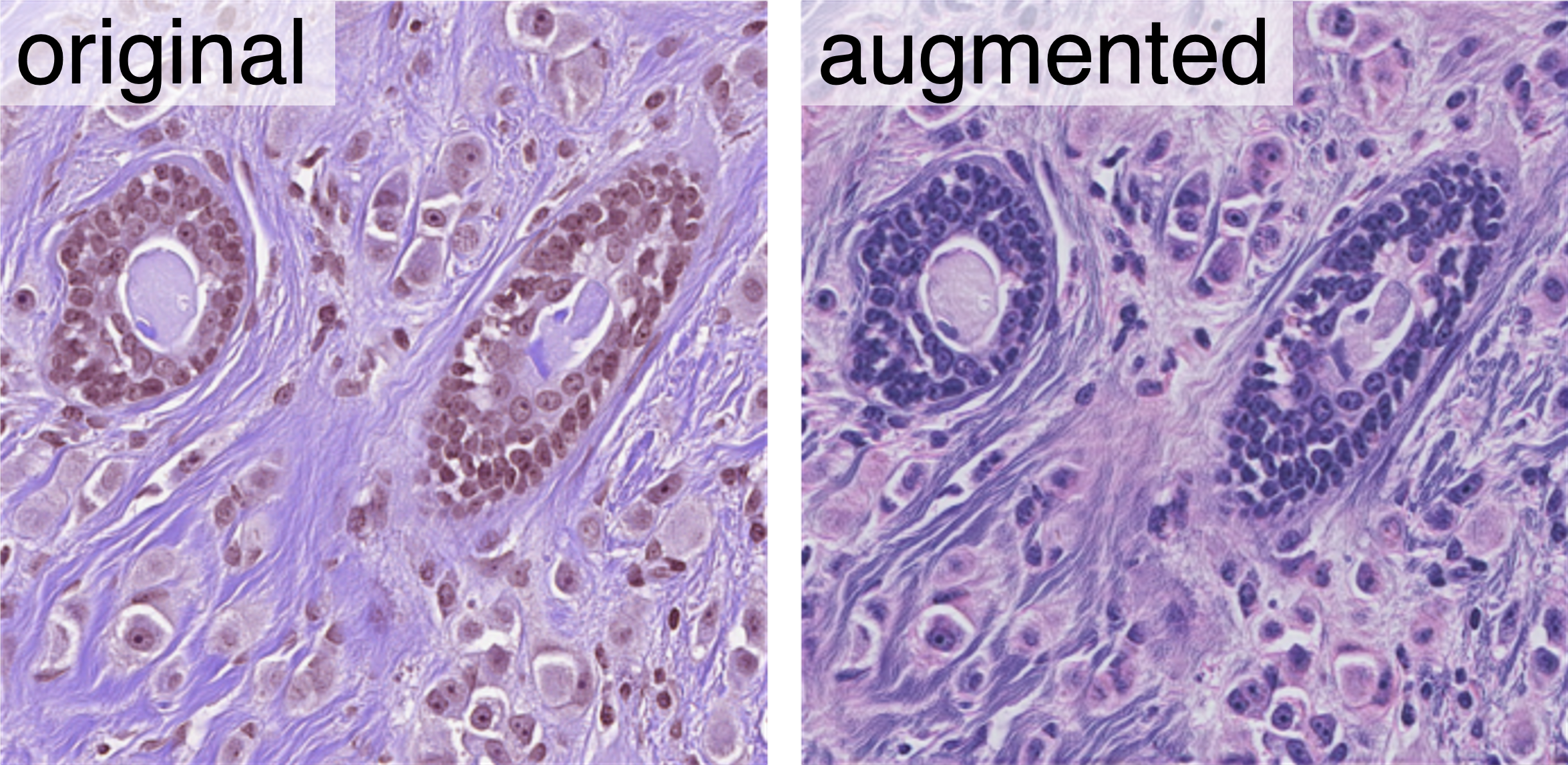}
}
\end{floatrow}
\end{figure}

We further check the quality of HistAuGAN by an expert pathologist on the structural similarity of original and augmented WSI tiles from the training set, i.e., the Camelyon17 dataset, and an unseen dataset of breast tissue \cite{mitosisdataset}. We define three levels of similarity: a) “High similarity”: a pathologist would find it difficult to distinguish the original tile from the augmented tile. b) “Moderate similarity”: some structural variations are observed, but do not affect pathological diagnosis. c) “Low similarity”: the augmentated tiles can not be used for diagnostic purposes. As shown in Table \ref{tab:expert}, most of the augmented images do not have a structural modification that affects diagnosis and over half of them can even fool an expert pathologist. It is worth mentioning that HistAuGAN is not trained on any of the breast cancer images but is still able to transfer its color in a structure-preserving manner as shown in Figure \ref{fig:breast} on a sample tile.

\section{Conclusion}
In summary, we propose a novel GAN-based technique, HistAuGAN, for color augmentation of histopathological images. Based on the disentangled representations of content and style, HistAuGAN is able to change the color appearance of an histological image while preserving its morphological structure. Moreover, HistAuGAN captures both intra-domain and inter-domain color variations. It is able to interpolate between domains and can therefore span a continuous color space covering a large variety of realistic stain colors. When applied as an augmentation technique, HistAuGAN yields a robust down-stream classifier that generalizes better to out-of-domain test sets than other color augmentations techniques and, therefore, renders additional stain normalization steps unnecessary.  
Finally, \mbox{HistAuGAN} can mitigate batch effects present in histopathological data which suggests that it is also able to cover domain shifts beyond color variations, such as noise and artifacts introduced in image compression. The code is publicly available at \url{https://github.com/sophiajw/HistAuGAN} together with a model of HistAuGAN trained on the five medical centers of the Camelyon17 dataset. 

%
%
%
\bibliographystyle{splncs04}
\bibliography{bibliography.bib}

\begin{thebibliography}{10}
\providecommand{\url}[1]{\texttt{#1}}
\providecommand{\urlprefix}{URL }
\providecommand{\doi}[1]{https://doi.org/#1}

\bibitem{Bandi2019-bo}
Bandi, P., Geessink, O., Manson, Q., Van~Dijk, M., Balkenhol, M., Hermsen, M.,
  Ehteshami~Bejnordi, B., Lee, B., Paeng, K., Zhong, A., Li, Q., Zanjani, F.G.,
  Zinger, S., Fukuta, K., Komura, D., Ovtcharov, V., Cheng, S., Zeng, S.,
  Thagaard, J., Dahl, A.B., Lin, H., Chen, H., Jacobsson, L., Hedlund, M.,
  Cetin, M., Halici, E., Jackson, H., Chen, R., Both, F., Franke, J.,
  Kusters-Vandevelde, H., Vreuls, W., Bult, P., van Ginneken, B., van~der Laak,
  J., Litjens, G.: From detection of individual metastases to classification of
  lymph node status at the patient level: The {CAMELYON17} challenge. IEEE
  Trans. Med. Imaging  \textbf{38}(2),  550--560 (Feb 2019)

\bibitem{Becht2018-hk}
Becht, E., McInnes, L., Healy, J., Dutertre, C.A., Kwok, I.W.H., Ng, L.G.,
  Ginhoux, F., Newell, E.W.: Dimensionality reduction for visualizing
  single-cell data using {UMAP}. Nat. Biotechnol.  (Dec 2018)

\bibitem{Bejnordi2016-eb}
Bejnordi, B.E., Litjens, G., Timofeeva, N., Otte-Holler, I., Homeyer, A.,
  Karssemeijer, N., van~der Laak, J.A.: Stain specific standardization of
  {Whole-Slide} histopathological images (2016)

\bibitem{De_Bel2021-ay}
de~Bel, T., Bokhorst, J.M., van~der Laak, J., Litjens, G.: Residual cyclegan
  for robust domain transformation of histopathological tissue slides. Med.
  Image Anal.  \textbf{70},  102004 (May 2021)

\bibitem{Chan2014-ns}
Chan, J.K.C.: The wonderful colors of the {Hematoxylin--Eosin} stain in
  diagnostic surgical pathology. Int. J. Surg. Pathol.  \textbf{22}(1),  12--32
  (Feb 2014)

\bibitem{hallucinatingGANs}
Cohen, J.P., Luck, M., Honari, S.: Distribution matching losses can hallucinate
  features in medical image translation. In: Frangi, A.F., Schnabel, J.A.,
  Davatzikos, C., Alberola{-}L{\'{o}}pez, C., Fichtinger, G. (eds.) Medical
  Image Computing and Computer Assisted Intervention - {MICCAI} 2018 - 21st
  International Conference, Granada, Spain, September 16-20, 2018, Proceedings,
  Part {I}. Lecture Notes in Computer Science, vol. 11070, pp. 529--536.
  Springer (2018). \doi{10.1007/978-3-030-00928-1\_60},
  \url{https://doi.org/10.1007/978-3-030-00928-1\_60}

\bibitem{resnet}
He, K., Zhang, X., Ren, S., Sun, J.: Deep residual learning for image
  recognition. In: Proceedings of the IEEE Conference on Computer Vision and
  Pattern Recognition (CVPR) (June 2016)

\bibitem{Lee2020-zw}
Lee, H.Y., Tseng, H.Y., Mao, Q., Huang, J.B., Lu, Y.D., Singh, M., Yang, M.H.:
  {DRIT} : Diverse {Image-to-Image} translation via disentangled
  representations (2020)

\bibitem{Liang2020-fl}
Liang, H., Plataniotis, K.N., Li, X.: Stain style transfer of histopathology
  images via {Structure-Preserved} generative learning. In: Machine Learning
  for Medical Image Reconstruction. pp. 153--162. Springer International
  Publishing (2020)

\bibitem{macenko2009method}
Macenko, M., Niethammer, M., Marron, J.S., Borland, D., Woosley, J.T., Guan,
  X., Schmitt, C., Thomas, N.E.: A method for normalizing histology slides for
  quantitative analysis. In: 2009 IEEE International Symposium on Biomedical
  Imaging: From Nano to Macro. pp. 1107--1110. IEEE (2009)

\bibitem{Nadeem2020-bn}
Nadeem, S., Hollmann, T., Tannenbaum, A.: Multimarginal wasserstein barycenter
  for stain normalization and augmentation. Med. Image Comput. Comput. Assist.
  Interv.  \textbf{12265},  362--371 (Oct 2020)

\bibitem{Reinhard2001-op}
Reinhard, E., Adhikhmin, M., Gooch, B., Shirley, P.: Color transfer between
  images. IEEE Comput. Graph. Appl.  \textbf{21}(5),  34--41 (Jul 2001)

\bibitem{mitosisdataset}
Roux, L.: Mitos-atypia-14 grand challenge.
  \url{https://mitos-atypia-14.grand-challenge.org/}, accessed: 2021-03-03

\bibitem{Shaban2019-sj}
Shaban, M.T., Tarek~Shaban, M., Baur, C., Navab, N., Albarqouni, S.: Staingan:
  Stain style transfer for digital histological images (2019)

\bibitem{Tellez2019-nt}
Tellez, D., Litjens, G., B{\'a}ndi, P., Bulten, W., Bokhorst, J.M., Ciompi, F.,
  van~der Laak, J.: Quantifying the effects of data augmentation and stain
  color normalization in convolutional neural networks for computational
  pathology. Med. Image Anal.  \textbf{58},  101544 (Dec 2019)

\bibitem{Tschuchnig2020-tj}
Tschuchnig, M.E., Oostingh, G.J., Gadermayr, M.: Generative adversarial
  networks in digital pathology: A survey on trends and future potential.
  Patterns (N Y)  \textbf{1}(6),  100089 (Sep 2020)

\bibitem{Vahadane2016-lu}
Vahadane, A., Peng, T., Sethi, A., Albarqouni, S., Wang, L., Baust, M.,
  Steiger, K., Schlitter, A.M., Esposito, I., Navab, N.: {Structure-Preserving}
  color normalization and sparse stain separation for histological images. IEEE
  Trans. Med. Imaging  \textbf{35}(8),  1962--1971 (Aug 2016)

\bibitem{randomerasing}
Zhong, Z., Zheng, L., Kang, G., Li, S., Yang, Y.: Random erasing data
  augmentation. Proceedings of the AAAI Conference on Artificial Intelligence
  \textbf{34}(07),  13001--13008 (Apr 2020). \doi{10.1609/aaai.v34i07.7000},
  \url{https://ojs.aaai.org/index.php/AAAI/article/view/7000}

\bibitem{Zhu2017-li}
Zhu, J.Y., Park, T., Isola, P., Efros, A.A.: Unpaired image-to-image
  translation using cycle-consistent adversarial networks. In: Proceedings of
  the {IEEE} international conference on computer vision. pp. 2223--2232 (2017)

\end{thebibliography}

\end{document}